\pdfoutput=1 
\documentclass[sigconf]{acmart}

\def\BibTeX{{\rm B\kern-.05em{\sc i\kern-.025em b}\kern-.08emT\kern-.1667em\lower.7ex\hbox{E}\kern-.125emX}}
    
\copyrightyear{2019}
\acmYear{2019} 
\setcopyright{iw3c2w3}
\acmConference[WWW '19 Companion]{Companion Proceedings of the 2019 World Wide Web Conference}{May 13--17, 2019}{SFO, CA, USA}
\acmBooktitle{Companion Proceedings of the 2019 World Wide Web Conference (WWW '19 Companion), May 13--17, 2019, San Francisco, CA, USA}
\acmPrice{}
\acmDOI{10.1145/3308560.3316456}
\acmISBN{978-1-4503-6675-5/19/05}

\acmSubmissionID{laweb07}

\usepackage[utf8]{inputenc}
\usepackage[T1]{fontenc}
\usepackage{longtable}
\usepackage{graphicx}
\usepackage{booktabs}
\usepackage{multirow}

\usepackage{color, colortbl}
\definecolor{Gray}{gray}{0.9}
\definecolor{LightCyan}{rgb}{0.88,1,1}
\definecolor{Violet}{RGB}{252, 243, 207}

\usepackage{url}

\begin{document}

%




\title[Global Reactions to the Cambridge Analytica Scandal]{Global Reactions to the Cambridge Analytica Scandal:  A Cross-Language Social Media Study}

\author{Felipe Gonz\'{a}lez}
\affiliation{%
  \institution{UTFSM}
  \city{Santiago}
  \country{Chile}
}
\email{felipe.gonzalezpi@usm.cl}

\author{Yihan Yu}
\affiliation{%
  \institution{University of Washington}
  \city{Seattle}
  \country{United States}
}
\email{yyu2016@uw.edu}

\author{Andrea Figueroa}
\affiliation{%
  \institution{UTFSM}
  \city{Valpara\'{i}so}
  \country{Chile}
}
\email{andrea.figueroa@usm.cl}

\author{Claudia L\'{o}pez}
\affiliation{%
  \institution{UTFSM}
  \city{Valpara\'{i}so}
  \country{Chile}
}
\email{claudia@inf.utfsm.cl} 

\author{Cecilia Aragon}
\affiliation{%
  \institution{University of Washington}
  \city{Seattle}
  \country{United States}
}
\email{aragon@uw.edu}

%

%
\begin{abstract}

Currently, there is a limited understanding of how data privacy concerns vary across the world. The Cambridge Analytica scandal triggered a wide-ranging discussion on social media about user data collection and use practices. We conducted a cross-language study of this online conversation to compare how people speaking different languages react to data privacy breaches. We collected tweets about the scandal written in Spanish and English between April and July 2018. We used the Meaning Extraction Method in both datasets to identify their main topics. They reveal a similar emphasis on Zuckerberg's hearing in the US Congress and the scandal's impact on political issues. However, our analysis also shows that while English speakers tend to attribute responsibilities to companies, Spanish speakers are more likely to connect them to people. These findings show the potential of cross-language comparisons of social media data to deepen the understanding of cultural differences in data privacy perspectives.   


\end{abstract}

%
%


%

%
\maketitle

\section{Introduction}

While there is evidence that concerns about privacy and its intricate relationship with users' decisions to use social media have been rising among Americans 
\cite{smith1996information}, 
much less is known about these perspectives across the world. So far, most of the literature about privacy concerns has been focused on the United States and even though recently research has begun to be carried out examining European privacy perspectives after the General Data Protection Regulation (GDPR) implementation, privacy research in international settings is still needed \cite{smith2011information}. A few survey studies of people from different pairs of countries have been conducted to address. This, and the results so far   indicate that data privacy perspectives vary significantly across countries. 
Belgians self-reported lower levels of concerns about sensitive information leakage than people from the United States. Harris et al. attributed this difference to the privacy laws of the respondents' countries of origin \cite{harris2003privacy, smith2001information}. 
A survey of North American 
and Turkish freshmen living in similar residence hall settings \cite{kaya2003cross} 
showed that Americans wished for more privacy in their 
hall rooms than Turkish students. 
In the context of e-commerce, another survey 
found that Italians tend to exhibit lower privacy concerns than Americans \cite{dinev2006privacy}. 

Unfortunately, relying solely on survey data for cross-cultural studies of data privacy has various limitations. 
Most of them focus only on 
two geographic regions 
and have limited sample size 
\cite{ur2013cross}. 
Additionally, most privacy surveys are only available 
in English and only a few of them have been translated to other 
languages 
\cite{ur2013cross}. 
Surveys of a multinational or global nature that can mitigate these limitations would be very costly, 
which makes it difficult to compare privacy attitudes 
more broadly. 

As the Cambridge Analytica scandal unfolded and people became aware that the personal data of 87 million Facebook users were exposed without their consent and used by Cambridge Analytica to support political campaigns 
\cite{lapaire2018content},
thousands of people in different parts of the world expressed on social media their reactions to and reflections on the scandal, its relationship to data privacy, and its broader implications. Indeed, a movement to delete Facebook accounts emerged and the \#deleteFacebook hashtag was trending for several days \cite{mirchandani2018delete}.

In this paper, we report on a study that observes Twitter activity about the Cambridge Analytica scandal in Spanish and English and proposes a methodology for cross-language comparison of social media text. We believe that our approach offers an alternative or complementary method to conduct studies on data privacy perspectives across speakers of different languages and may provide a roadmap for future cross-cultural research.  
As Twitter allows people to express themselves freely and spontaneously and in different languages, it enables a unique opportunity to analyze multi-language large-scale data. These characteristics allow researchers to address some of the limitations of the survey-based methods described above, such as those related to language and sample size \cite{wang2017exploring}. Our premise is that written communication can be a ``window into culture and an external reflection of cultural values'' \cite{choi2016cross}; therefore, what people write about the scandal in their own languages can reveal differences and similarities in data privacy concerns across users worldwide.

We summarize prior work on cross-language comparisons of social media in Section \ref{sec:relatedwork}. Section \ref{sec:question} introduces our research question, Section \ref{sec:method} details our research method, Section \ref{sec:results} reports on our findings, and Section \ref{sec:discusion} offers a discussion of our results, its limitations, and future work. Finally, Section \ref{sec:conclusions} provides our conclusions.

\section{Cross-language comparisons of user-generated written content}
\label{sec:relatedwork}

Prior research has used social media text in different languages to make comparisons among people who speak these languages. An analysis of more than 62 million tweets compared the top 10 most common languages regarding the use of features such as: URLs, hashtags, mentions, replies and re-tweets \cite{hong2011language}. The findings show that German-speaking users tend to include more URLs and hashtags in their tweets than other users, while 
Korean-speaking users are prone to reply to each other more often than speakers of other languages. Hong et al. argue that users of different languages use Twitter for different purposes. The German community often uses this platform for information sharing, while the Korean community employs it for conversational purposes. Another study analyzes tweets written by Americans in English and Japanese in their official language \cite{acar2013culture}. As opposed to Americans, Japanese people tweet more self-related messages and more messages about TV programs. In turn, Americans tweet more about their peers, sports and news. 

Previous work has also explored how user-generated content can reveal different views of the same issues among people who write in different languages. 
The Meaning Extraction Method (MEM) was applied to compare posts from depression-related forums in Spanish and English \cite{ramirez2008psychology}. MEM is used to discover the main topics in a corpus. A comparison of the resulting topics 
shows that English posts tend to use words that are more concrete and descriptive and the main topics are 
related to medicinal questions and concerns. Spanish posts use relatively more emotional words and the main topics are 
associated with sharing and disclosing information about relational concerns. Hecht and Gergle used the Explicit Semantic Analysis (ESA) algorithm to analyze pairs of terms from ten different Wikipedia language editions \cite{hecht2010tower}. ESA indicates a score of semantic relatedness between two concepts. The findings reveal that ``even when two language editions cover the same concept, they may describe that concept differently'' \cite{hecht2010tower}. For example, consider the pair ``Germany'' / ``Saxony-Anhalt'' (a state of Germany). In most languages this pair receives a high ESA score, but the algorithm detects no relation at all in Italian and Danish. This occurs because there are no articles that mention ``Germany'' and ``Saxony-Anhalt'' together in these two languages. 
Analyses of semantic networks and the salience of semantic concepts in articles about China in the Chinese and English versions of Wikipedia found dissimilarities in the semantic content of these two versions \cite{jiang2017mapping}. Articles in the Chinese version are framed from the perspectives of respecting authority, emphasizing harmony and patriotism. Articles in English are written from the perspective of Western-societies' core value of democracy. The English version contains critical attitudes toward the authority of the Chinese government and the Communist Party in terms of human rights and territorial dispute. According to Jiang et al. cultures, values, interests, situations and emotions of different language groups can explain these dissimilarities. 
The latter studies provide evidence in support of what is known as the Sapir‐Whorf hypothesis, which indicates that the structure of anyone's native language influences the world-views she or he will acquire as she or he learns the language \cite{kay1984sapir}. Thus, speakers of different languages could think, perceive reality and organize the world around them  in different ways \cite{hussein2012sapir}. 
Inspired by this hypothesis, we seek to study whether people who speak a different language hold different views of a data privacy scandal, such as the Cambridge Analytica case.

\section{Research Question}
\label{sec:question}

Given the relative lack of data privacy research in international settings \cite{smith2011information}, our project aims to investigate the potential of social media text written in different languages as a source to compare data privacy views worldwide. Beyond the Sapir‐Whorf hypothesis (explained in Section \ref{sec:relatedwork}), prior privacy research has argued that language and country of residence can relate to diverse perspectives on privacy. Smith et al. \cite{smith2011information} noted that 
``many languages, including those in European countries (e.g Russian, French, Italian), do not have a word for privacy and have adopted the English word''. Belanger and Crossler argued that ``individuals from different countries can be expected to have different cultures, values and laws, which may result in differences in their perceptions of information privacy and its impacts'' \cite{belanger2011privacy}. 

As the Cambridge Analytica scandal sparked worldwide conversations (in diverse languages) on Twitter about this particular misuse of user data, these online public communications are useful sources to contrast views on the scandal itself, its relation to data privacy, and its implications. To start addressing our ultimate research goal, this paper focuses on the following research question: ``Which are the shared and unique topics that emerge from the Twitter activity in Spanish and English about the Cambridge Analytica data misuse scandal?'' 

\section{Data and methods}
\label{sec:method}

To answer our research question, we used Tweepy\footnote{http://www.tweepy.org/}, a Python library for accessing to the standard realtime streaming Twitter API. Using this library we were able  to capture tweets that include hashtags or keywords 
related to the Cambridge Analytica scandal or data privacy, such as: ``\#CambridgeAnalytica'', ``\#DeleteFacebook'', ``Zuckerberg'' and ``Facebook privacy''. The  standard realtime streaming Twitter API returns 
a random sample of all public tweets that match the search keywords. 
We collected tweets written in Spanish and English between April 1st and July 10th, 2018. 
Overall, we collected more than 7.4 million tweets written in English and 
more than $470,000$ tweets in Spanish (see Table \ref{Table:NumberOfTweets}). The English tweets were generated by about 1.8 million unique Twitter accounts while the Spanish tweets were produced by approximately $220,000$ users. The difference between the number of tweets and users collected in English and Spanish suggests that English-speaking Twitter users tweeted more about this scandal using the selected keywords than Spanish-speaking users, although this may be explained by the greater volume of English tweets overall\footnote{https://www.statista.com/statistics/267129/most-used-languages-on-twitter/}.

We cleaned our dataset in two ways. First, we removed all retweets to focus our study on original user opinions and avoid analyzing duplicates. This step downsized both datasets to approximately 20\% of their original sizes. Second, we attempted to eliminate tweets generated by automated accounts so our study could indeed reflect people's opinions. Unfortunately, there is not yet an infallible mechanism to detect bots' activity on Twitter. We chose to use Botometer to identify potential bots \cite{davis2016botornot}. Botometer implements a machine learning algorithm that has achieved high accuracy ($0.94$) in detecting both simple and sophisticated bots in prior work \cite{varol2017online}. The algorithm has been trained to detect bots by analyzing Twitter accounts' metadata, their contacts' metadata, tweets' content and sentiment, network patterns, and activity time series. 
The result is a score that is based on how likely it is to be a bot. The score ranges from 0 to 1, where lower scores indicate that the account behaves like a human and higher scores signal bot-like behavior. Unfortunately, there is not yet agreement on a threshold that can reliably distinguish bots from humans. To define thresholds for our two datasets, 
we used the Ckmeans \cite{wang2011ckmeans} algorithm to cluster the Botometer scores in each dataset into five groups, with the first cluster including the accounts with the lowest Botometer scores (more human-like) and the fifth group the users with the highest scores (more bot-like). We reasoned that the fourth and fifth clusters in each dataset were least likely to contain humans; therefore, we excluded them from our analysis.
Given that the Botometer analysis
is time-consuming, we analyzed only a sample of users. In this study, we focused on the 
users 
who contributed the highest number of tweets in our datasets. In the future we plan to analyze users who contribute less. 
Thus, we were able to classify $19,478$ accounts in the Spanish dataset (40.6\%) and 74,021 (12.9\%) accounts in the English dataset. 
Accounts with a Botometer score higher than $0.4745$ were labeled as bots in the Spanish dataset. Those with a score higher than $0.4849$ were considered bots in the English dataset. These users and their tweets were removed from our analysis. 
As a result, our final Spanish dataset includes $15,531$ users who tweeted $50,559$ times about the Cambridge Analytica scandal. The English dataset comprises $60,491$ accounts that generated $446,462$ tweets about it. 
Table \ref{Table:NumberOfTweets} details these figures. 
\begin{table}[htb]
\caption{Size of the Spanish and English datasets before and after data cleaning }

\resizebox{\columnwidth}{!}{%
\begin{tabular}{|l|rr|rr|}
\hline
Dataset             & \multicolumn{2}{c|}{Spanish}  & \multicolumn{2}{c|}{English}              \\ \hline
                    & \#Tweets      & \#Users       & \#Tweets      & \#Users   \\ \hline
Total               & 472,363     & 222,352     & 7,476,988   & 1,846,542   \\

\end{tabular}
}

\end{table}

\begin{table}[htb]
\caption{Size of the Spanish and English datasets before and after data cleaning }
\label{Table:NumberOfTweets}
\resizebox{\columnwidth}{!}{%
\begin{tabular}{|l|rr|rr|}
\hline
Dataset             & \multicolumn{2}{c|}{Spanish}  & \multicolumn{2}{c|}{English}              \\ \hline
                    & \#Tweets      & \#Users       & \#Tweets      & \#Users   \\ \hline
Total               & 472,363     & 222,352     & 7,476,988   & 1,846,542   \\
Without retweets    & 106,656     & 47,951      & 1,572,371   & 574,452       \\
\hline
Most active         & 70,393      & 19,478      & 741,694     & 74,021      \\ 
Humans              & 50,559      & 15,531      & 446,462     & 60,491      \\ \hline

\end{tabular}
}
\end{table}

\begin{table}[htb]
\caption{Size of the Spanish and English datasets before and after data cleaning }
\label{Table:NumberOfTweets}
\resizebox{\columnwidth}{!}{%
\begin{tabular}{@{\extracolsep{6pt}}llrrrr}
\toprule
Dataset             & \multicolumn{2}{c}{Spanish}  & \multicolumn{2}{c}{English}              \\ 
    \cmidrule{2-3} 
    \cmidrule{4-5} 
                    & \#Tweets      & \#Users       & \#Tweets      & \#Users   \\
\midrule
Total               & 472,363     & 222,352     & 7,476,988   & 1,846,542   \\
Without retweets    & 106,656     & 47,951      & 1,572,371   & 574,452       \\
Most active users         & 70,393      & 19,478      & 741,694     & 74,021      \\ 
\textbf{Humans}              & \textbf{50,559}      & \textbf{15,531}      & \textbf{446,462}     & \textbf{60,491}      \\ 
\bottomrule
\end{tabular}
}
\end{table}

\begin{table}[htb]
\caption{Size of the Spanish and English datasets collected}
\label{Table:NumberOfTweets}
\resizebox{\columnwidth}{!}{%
\begin{tabular}{@{\extracolsep{6pt}}llrrrr}
\toprule
Dataset             & \multicolumn{2}{c}{Spanish}  & \multicolumn{2}{c}{English}              \\ 
    \cmidrule{2-3} 
    \cmidrule{4-5} 
                    & \#Tweets      & \#Users       & \#Tweets      & \#Users   \\
\midrule
Total               & 472,363     & 222,352     & 7,476,988   & 1,846,542   \\
\bottomrule
\end{tabular}
}
\end{table}

To identify key topics in the resulting Spanish and English datasets, we used the Meaning Extraction Method (MEM) \cite{chung2008revealing}. MEM is a topic modeling technique that can  infer ``what words are being used together, essentially resulting in a dictionary of word-to-category mappings from a collection of texts''\cite{boyd2017psychological}. After applying principal component analysis over this dictionary it is possible to identify words that can be grouped into themes or topics. 
This method has been identified as well-suited for cross-cultural and cross-language research \cite{boyd2017psychological}. MEM has been used to find themes in different contexts, such as
: mental health \cite{wolf2010inpatient,
ramirez2008psychology}; personality \cite{boyd2015did,griffin2018online} and values 
\cite{wilson2016disentangling}.

We employed the Meaning Extraction Helper (MEH) software \cite{boyd2014meh}, tool that can automate the majority of the MEM process \cite{boyd2017psychological}. 
The software contains a default list of Spanish and English stopwords; all these words were removed. Also, the tool allowed us to apply text segmentation by whitespace, conduct lemmatization and run Twitter-aware tokenization. To assist in lemmatization tasks, a conversion list was used to fix common misspellings (e.g ``hieght'' to ``height'') and convert ``textisms'' (e.g, ``bf'' to ``boyfriend''). No stemming algorithm was used. We computed the frequency of each unigram as the percentage of tweets that contain it. The 300 most frequent unigrams 
were 
kept. We obtained a csv file with values of 1 and 0 indicating the corresponding unigrams' presence or absence, respectively, for each tweet.

Principal component factor analysis (PCA) was run over the MEH results. 
PCA was performed 
with varimax rotation 
to ensure 
that all resulting components 
are independent 
from each other. 
We conducted PCA 
with 5, 8, 11, 30 and 100 components. 
In both the Spanish and English datasets, 11 
components gave the best results, with  
fit based upon diagonal scores of $0.55$ and $0.9$, respectively. 
This metric is a goodness of fit statistic, where values closer to 1 indicate better fit. 
The selected components accounted for $9\%$ of the total variance of the Spanish data and $13\%$ of the total variance of the English data. 
 
To obtain the most representative words of each resulting component, we selected the words with factor loadings above $0.1$, as recommended in \cite{boyd2017psychological}. 
The words were sorted according to their contribution 
to the component. Additionally, a python script was used to identify the top-30 tweets most related to each component. Using the most representative words and tweets by component, two authors examined and conceptualized the theme represented by each component, assigned a representative name and determined its relevance to our research question.

Finally, we used GeoNames API\footnote{http://www.geonames.org/} to geo-locate all human-like accounts (see Table \ref{Table:NumberOfTweets}). 
Table \ref{Table:Geolocated} reports the proportion of tweets and users by the top-10 countries in each of our datasets. Most accounts could not be geo-located. The remaining accounts reveal that Spain and the US account for the majority of tweets and users in the Spanish and English datasets, respectively. 

\begin{table}[htb]
\caption{Top ten most frequent user location in the English and Spanish datasets}
\label{Table:Geolocated}
\resizebox{\columnwidth}{!}{%
\begin{tabular}{@{}|c|c|c|c|c|c|@{}}
\hline
\multicolumn{3}{|c|}{Spanish}        & \multicolumn{3}{c|}{English}          \\ \hline
Country       & \% tweets & \% users & Country        & \% tweets & \% users \\ \hline
not found   & 43.5\%      & 46.05\%    & not found   & 41.6\%    & 44.9\%   \\ \hline
Spain       & 17.5\%      & 16.43\%    & U.S         & 33.6\%    & 31.1\%   \\ \hline
Mexico      & 10.3\%      & 9.90\%     & U.K         & 6.2\%     & 6.9\%    \\ \hline
Venezuela   & 5.6\%       & 3.56\%     & India       & 3.3\%     & 3.0\%    \\ \hline
Argentina   & 5.4\%       & 5.41\%     & Canada      & 2.4\%     & 2.3\%    \\ \hline
Colombia    & 2.8\%       & 2.98\%     & Australia   & 1.1\%     & 1.4\%    \\ \hline
U.S         & 2.3\%       & 2.40\%     & France      & 1.0\%     & 0.6\%    \\ \hline
Chile       & 2.0\%       & 2.35\%     & Germany     & 0.9\%     & 0.7\%    \\ \hline
Peru        & 1.3\%       & 1.41\%     & U.A.E       & 0.6\%     & 0.3\%    \\ \hline
Ecuador     & 1.2\%       & 1.25\%     & Netherlands & 0.5\%     & 0.4\%    \\ \hline
Brazil      & 0.9\%       & 0.34\%     & Ireland     & 0.4\%     & 0.4\%    \\ \hline
\end{tabular}
}
\end{table}



\section{Results}
\label{sec:results}

The words that clustered together to form coherent themes in the English and Spanish corpora are available online\footnote{https://github.com/gonzalezf/LA-WEB-Paper}. Tables \ref{Table:SpanishTerms} and \ref{Table:EnglishTerms} report the seven words with the highest loadings by component in the English and Spanish datasets, respectively. The tables also show the proportion explained (PE) by each of them according to the factor analysis. This number is proportional to the number of tweets associated with each theme. Table \ref{Table:ThemesFound} presents the key themes in Twitter activity about the Cambridge Analytica scandal in Spanish and English. The themes are ordered according to their relevance to our research question.

\begin{table*}[]
\caption{Terms by themes in the Spanish dataset}
\label{Table:SpanishTerms}
\begin{tabular}{|c|c|c|c|c|c|c|c|c|}
\hline
\multirow{2}{*}{\#ID} & \multirow{2}{*}{PE (\%)} & \multicolumn{7}{c|}{Words with the highest loadings by component}                                          \\ \cline{3-9} 
                      &                          & word1            & word2            & word3      & word4      & word5          & word6    & word7          \\ \hline
S1                    & 13\%                   & animals       & lottery      & predict    & live    & program        & listen   & win            \\ \hline
S2                    & 11\%                   & cambridge        & analytica        & million    & user       & scandal        & data     & affect         \\ \hline
S3                    & 10\%                   & congress         & error            & zuckerberg & sorry      & mark           & usa     & senate         \\ \hline
S4                    & 9\%                    & rgpd             & protection       & gdpr       & data       & regulation     & privacy  & dataprotection \\ \hline
S5                    & 9\%                    & marketing        & digitalmarketing & analytic   & publicity  & digital        & google   & youtube        \\ \hline
S6                    & 9\%                    & press             & like             & followers  & platform & follow-us       & come     & digital            \\ \hline
S7                    & 9\%                    & red              & social           & twitter    & instagram  & youtube        & facebook & follow-us             \\ \hline
S8                    & 8\%                    & parliament       & european         & zuckerberg & mark       & ask            & hearing  & sorry          \\ \hline
S9                    & 8\%                    & stop             & ask              & join       & people     & red            & create   & senate         \\ \hline
S10                   & 8\%                    & communitymanager & socialmedia          & blog       & socialnetworks & news     & work & facebook           \\ \hline
S11                   & 7\%                    & message          & messenger        & user       & send       & million        & remove   & delete  \\ \hline
\end{tabular}

\end{table*}

\begin{table*}[]
\caption{Terms by themes in the English dataset}
\label{Table:EnglishTerms}
\begin{tabular}{|c|c|c|c|c|c|c|c|c|}
\hline
\multirow{2}{*}{\#ID} & \multirow{2}{*}{\begin{tabular}[c]{@{}c@{}}PE (\%)\end{tabular}} & \multicolumn{7}{c|}{Words with the highest loadings by component}                                                     \\ \cline{3-9} 
                      &                                                                                       & word1        & word2     & word3                  & word4                 & word5         & word6        & word7      \\ \hline
E1                    & 24\%                                                                                & newyorkcity  & newyork   & nyc                    & ny                    & career        & code         & hire       \\ \hline
E2                    & 12\%                                                                                 & rsi          & btc       & signal                 & min                   & eth           & bitcoin      & crypto     \\ \hline
E3                    & 10\%                                                                                 & machinelearn & deeplearn & artificialintelligence & ml                    & robotic       & dl           & ai         \\ \hline
E4                    & 10\%                                                                                & chatbot      & infosec   & databreach             & hack                  & cybersecurity & crypto       & blockchain \\ \hline
E5                    & 9\%                                                                                 & iot          & iiot      & smartcity              & digitaltransformation & innovation    & infographic  & startup    \\ \hline
E6                    & 7\%                                                                                 & cambridge    & analytica & election               & campaign              & voter         & vote         & brexit     \\ \hline
E7                    & 7\%                                                                                 & data         & user      & facebook               & privacy               & access        & personal     & law        \\ \hline
E8                    & 6\%                                                                                 & zuckerberg   & mark      & testify                & ceo                   & congress      & committee    & senate     \\ \hline
E9                    & 6\%                                                                                 & trump        & president & donald                 & white                 & house         & democrat     & america    \\ \hline
E10                   & 5\%                                                                                 & social       & media     & twitter                & instagram             & censorship    & conservative & facebook   \\ \hline
E11                   & 5\%                                                                                 & late         & daily     & thanks                 & bigdata               & remove        & social       & ai         \\ \hline
\end{tabular}

\end{table*}

\begin{table*}[]
\caption{Themes in the Spanish and English datasets}
\label{Table:ThemesFound}
\begin{tabular}{|p{0.36\textwidth}|c|p{0.36\textwidth}|c|}
\hline
\multicolumn{2}{|c|}{Spanish}                                   & \multicolumn{2}{c|}{English}                                    \\ \hline
\multicolumn{1}{|c|}{Theme}                      & \#ID & \multicolumn{1}{c|}{Theme}                       & \#ID \\ \hline
Cambridge Analytica's impact on political issues & S2            & Cambridge Analytica's impact on political issues & E6           \\
Mark Zuckerberg's Senate hearing in the US       & S3            & Mark Zuckerberg's Senate hearing in the US       & E8           \\
General Data Protection Regulation               & S4            & General Data Protection Regulation               & E7           \\
Zuckerberg in front of the European Parliament   & S8            & Facts and opinions about Donald Trump            & E9           \\
Opinions about Mark Zuckerberg                   & S9            & Stop censorship on social media                  & E10           \\
Facebook deletes Zuckerberg's private messages   & S11            & Cryptocurrencies                                 & E2           \\
Digital marketing                                & S5            & Artificial intelligence                          & E3           \\
Promoting subscription to social platforms         & S7            & Blockchain                                       & E4           \\
Promoting likes in social platforms              & S6            & Internet of things                               & E5           \\
Lottery results                                  & S1           & News about privacy on social media               & E11           \\
News and random facts                            & S10           & Hiring tech jobs in New York                     & E1           \\ \hline
\end{tabular}%
\end{table*}

Three key themes emerge in both languages. Spanish and English speakers talk about:
\begin{itemize}
    \item ``Cambridge Analytica's impact on political issues''
    \item ``Mark Zuckerberg's Senate hearing in the USA,'' and
    \item ``General Data Protection Regulation.''
\end{itemize} 

However, differences appear in how these themes are articulated. In regard to the first topic, the scandal's connection to Russia was much less relevant for Spanish speakers than for English speakers. English tweets focus on how Russia might have used Cambridge Analytica to intervene in the 2016 US elections and the UK Brexit campaign. For example, this component includes the following tweet: \textit{``@ianbremmer @billmaher @RealTimers You want to know how Brexit happened and Trump got to win? Cambridge Analytica, Bannon, Mercers and the Bot Farms in Russia. They started testing MAGA, Build the Wall, Lock Her Up, Anti-Muslim sentiment, Anti-Immigrant propaganda. Putin worked hard at it since 2012''}. 
On the other hand, the token \textit{Russia} does not appear as a representative word of this theme in the Spanish dataset. Instead, Spanish tweets are centered on Cambridge Analytica's closure as a result of the scandal. This behavior could be explained by the users' country of residence and its closeness to a salient political issue related to Cambridge Analytica. The US accounts for the largest share of users who report their location in our English dataset (see Table \ref{Table:Geolocated}). It is well known that many people from this country have apprehensions about Russia since the Cold War. Furthermore, recent research has found evidence that Americans express ``continued mistrust of Russia and a majority think Russia tried to interfere in the 2016 election'' \cite{smeltz2018public}. We believe that this is a plausible reason to explain this difference between the English and Spanish tweets.

While both datasets include a topic about ``Mark Zuckerberg's Senate hearing in the US'', their tweets' verb tenses differ. English-speaking users tend to tweet about this topic in future or present tense. These tweets reflect either certain level of anticipation of the event or live reports on how the event was unfolding. An example of these tweets is: 
\textit{``Facebook CEO Mark Zuckerberg will testify in front of a joint hearing of the Senate Judiciary and Commerce Committees today. Senators will demand answers from Zuckerberg about Facebooks failure to protect up to 87 million users' private information https://t.co/9qpdqxjM7F https://t.co/aLlZtMMVCC''}. On the other hand, Spanish-speaking users are more likely to discuss this issue in past tense. They comment on sentences that Zuckerberg said during the hearing, putting special emphasis on the moment when he assumed responsibility for what has happened. For instance, translated Spanish tweets state: \textit{``From \#Whoknows to \#ThroughMyFault, the change of attitude of \#MarkZuckerberg. `It was my mistake, and I'm sorry, I started \#Facebook and I'm responsible for what happens here': Mark Zuckerberg before the Congress of \#EEUU https://t.co/UX2QwT7pFw''}, and \textit{``Zuckerberg takes full blame for the abuse of Cambridge Analytica before the US Senate: `It was my mistake, and I'm sorry' https://t.co/VjH9ocdLcB''}. 
The difference in tenses (English future/present and Spanish past) may be explained by a delay in news reporting (volume of the Spanish tweets in our dataset relating to this particular topic tended to peak about 24 hours after the peak occurred in the English tweets) and translation to another language as the event occurred in an English-speaking country. 

In the case of the ``General Data Protection Regulation'' (GDPR) theme, a tendency to attribute responsibility to companies for not attempting to comply with the GDPR was present in the English data, but not in the Spanish tweets. The English dataset includes accusatory tweets to companies such as Facebook and Google for trying to dodge GDPR rules, e.g. \textit{``Facebook and Google are pushing users to share private information by offering invasive and limited default options despite new EU data protection laws aimed at giving users more control and choice https://t.co/x0srYwZeUz''}, \textit{``If a new European personal data regulation (aka \#GDPR) went into effect tomorrow, almost 1.9bn \#Facebook users around the world would be protected by it. The online social network is making changes that ensure the number will be much smaller.https://t.co/UXonA0mTCs \#privacy https://t.co/gqFaMYq9Su''}, and \textit{``Facebook moves billions of international user accounts to California to avoid European privacy law https://t.co/nC2ccf2ClR''}. On the other hand, Spanish-speaking users tweet about GDPR to describe it and inform
local companies how to prepare for it. Some translated tweets belonging to this component are: \textit{`` General \#Data Protection Regulation (\#GPDR) is a new \#law of the European Union and it will enter into force on May 25th. Do you know what it is? Do you know its characteristics? Is your \#company ready? Get up to date by clicking on the following link! https://t.co/AeKPR4w6Q9 https://t.co/rdlzbrAU1n''}, and \textit{``On May 25th, the General Protection \#Data Regulation (GDPR) of the EU, one of the most modern regulations regarding personal data use by companies and institutions, begins to be applied. There will be some repercussions in \#Chile. https://t.co/XNbmALUcdD''}

Three other emerging themes are unique to Spanish speakers and are relevant to our research question as they all refer to the founder of Facebook and its role in different aspects of the scandal. These topics are:
\begin{itemize}
    \item ``Zuckerberg in front of the European Parliament''
    \item ``Opinions about Mark Zuckerberg,'' and
    \item ``Facebook deletes Mark Zuckerberg's private messages.''
\end{itemize}   

The first topic 
focused on Zuckerberg's laments for the situation, e.g. \textit{``Mark Zuckerberg apologized to the European Parliament for the data breach. The founder of Facebook acknowledged on Tuesday that the tools of the social network were used `to do harm'. https://t.co/aeDUVViUhq''} We note that this theme only emerges in the Spanish tweets.
Again, we attribute this distinction to the users' country of residence. Both datasets include users whose self-reported location is in Europe; however, they are the majority only in the Spanish dataset where Spain is associated with more tweets than any other country (see Table \ref{Table:Geolocated}). Thus, the hearing in the EU Parliament is much more prominent in the Spanish dataset. 

The second topic contains supporting and accusatory tweets for Mark Zuckerberg in relation to the Cambridge Analytica scandal. This theme includes tweets such as the following: \textit{``what they do not forgive to Zuckerberg is that Facebook has unwittingly helped Trump triumph (which I do not think), because when Facebook censored pages of right-wing groups for nothing (the same Mark Zuckerberg has said that he worries about the leftist prejudice of his staff), there they said nothing.''} 

The third theme groups together tweets about the option to delete private direct messages on Facebook. This functionality apparently was available to Mark  Zuckerberg in the past while no other user could use it. Spanish speakers tweet about the special privileges of Zuckerberg as a Facebook user and about the possibility that such functionality could become available to everyone. Example translated tweets are: \textit{``Although you can not delete your messages sent from another person's inbox in Messenger, Facebook has done so in the case of Mark Zuckerberg and other company executives. https://t.co/PkJByaesdT''}, and \textit{``It is clear that Mark Zuckerberg is the only one who can have the God Mode of Facebook because only he can erase messages sent through Facebook Messenger, a feature that will be released for everyone after this scandal.''} 

Two themes are unique to English and are considered slightly relevant to our research question. They are conceptualized as: 
\begin{itemize}
    \item ``Stop censorship on social media,'' and
    \item  ``News about privacy on social media.''
\end{itemize}  

First, English-speaking users protest about censorship on social media companies such as Facebook, Twitter, Instagram and Google. They demand freedom of speech. This component includes the following tweet: 
\textit{``@facebook Social media censorship is tearing out our tongues! Is Facebook a neutral public forum? STOP CENSORSHIP! Demand an \#InternetBillOfRights \#IBOR  \#MAGA \#InvasionOfPrivacy \#HumanRights \#DeleteFacebook \#Censorship \#Twitter  \#Censorship \#Google  \#FaceBook \#Instagram  @realDonaldTrump 
''}. The next topic includes informative tweets (usually from electronic newspapers) that relate 
to social media and privacy, e.g. \textit{``The latest The GDPR and Data Protection Daily! https://t.co/GZBscXkxPl Thanks to @simpledatainc @Colt\_Technology @content\_app \#bigdata''}. 

Other key themes arise from both datasets, but we consider them not sufficiently relevant to our research question. In the Spanish dataset, these themes are: ``Digital marketing'', ``Promoting likes in social platforms'', ``Promoting subscription to social platforms'', ``Lottery results'', and ``News and random facts''.
In English, there are the following topics: ``Cryptocurrencies'', ``Artificial Intelligence'', ``Blockchain'', ``Internet of Things'', ``Facts and opinions about Donald Trump,'' and ``Hiring tech jobs in New York.''


\section{Discussion}
\label{sec:discusion}

\begin{table*}[]
\centering
\caption{Themes in the Spanish and English datasets}
\label{Table:ThemesFound}
\resizebox{\textwidth}{!}{
\begin{tabular}{ll}
\toprule
\multicolumn{1}{c}{Spanish}                                   & \multicolumn{1}{c}{English}                                    \\ \midrule
\rowcolor{LightCyan}
Cambridge Analytica's impact on political issues           & Cambridge Analytica's impact on political issues          \\
\rowcolor{LightCyan}
Mark Zuckerberg's Senate hearing in the US                   & Mark Zuckerberg's Senate hearing in the US                  \\
\rowcolor{LightCyan}

General Data Protection Regulation                           & General Data Protection Regulation                         \\
\rowcolor{Violet}

Zuckerberg in front of the European Parliament               &   Stop censorship on social media               \\
\rowcolor{Violet}
Opinions about Mark Zuckerberg                               &  News about privacy on social media                                                   \\
\cellcolor{Violet}  Facebook deletes Zuckerberg's private messages               & \cellcolor{Gray}Facts and opinions about Donald Trump                                             \\
\rowcolor{Gray}
Digital marketing                                           &    Cryptocurrencies                                  \\
\rowcolor{Gray}
Promoting subscription to social platforms                     &   Artificial intelligence                                             \\
\rowcolor{Gray}
Promoting likes in social platforms                          & Blockchain                                         \\
\rowcolor{Gray}
Lottery results                                             &  Internet of things\\
\rowcolor{Gray}
News and random facts                                       & Hiring tech jobs in New York                                \\ \bottomrule
\end{tabular}%
}
\end{table*}

Our study of tweets in Spanish and English about the Cambridge Analytica data misuse scandal allowed us to conduct a cross-language comparison of their main emerging themes. Out of eleven topics, three are common to both languages. However, they present meaningful differences in their articulation. Three other relevant themes are unique to the Spanish data and two relevant topics only arise from the English tweets. These findings show the potential of our algorithm-based approach for cross-language comparisons of tweets to identify similarities and (nuanced) differences on privacy-related views across people from different countries. 

We discuss here two underlying patterns that could explain the distinctions we found between Spanish and English speakers. 

First, we observe a tendency for a local perspective to appear in tweets about this data privacy scandal.  
Tweets in English provide elaborate rationales behind the Cambridge Analytica's impact on political issues. They often relate them to Russia's actions. This is a common argument among Americans \cite{smeltz2018public}, who are also the most active contributors of tweets in our English dataset. This kind of rationale is much less visible in tweets in the Spanish dataset, where most users are not located in the US. Furthermore, there is a large component of tweets about Mark Zuckerberg's hearing in front of the US Senate in both languages; however, only the Spanish corpus has a theme related to Zuckerberg's audience in the European parliament. Again, we relate this finding to the users' country of residence. Our Spanish dataset consists primarily of European users. This could explain why this event became more salient only in this language. Another finding that can be explained by a local perspective is the demand for freedom of speech in the English tweets. This theme does not emerge at all in the Spanish data. Free speech is a central value in the US \cite{alvarez2018free}, but it is not as prominent in Spanish-speaking countries. 


Second, English tweets often attribute responsibilities to governments and organizations while Spanish tweets tend to question individual actions. For example, English-speaking users tend to tweet accusatory statements about big companies such as Google and Facebook dodging GDPR. On the other hand, these large companies are very rarely questioned about their acts in the Spanish data. Instead, it is possible to find many tweets that support Mark Zuckerberg and blame Facebook users for being irresponsible by failing to read Facebook's terms of service and neglecting to protect their personal data. 

Research on cross-cultural differences could provide a framework to further explore the second pattern. This research argues that national cultures could be characterized by their scores on a small number of dimensions \cite{leidner2006review}. Hofstede's cultural dimensions \cite{hofstede1983national} have been widely used to study the relationship between people's culture and technology \cite{leidner2006review}. 
We hypothesize that one of these dimensions, ``power distance'', could explain differences between Spanish and English speakers on responsibility attribution. This dimension is defined as: ``the extent to which the less powerful members of institutions and organizations within a country expect and accept that power is distributed unequally'' \cite{hofstede2001culture}. In countries with a high level of power distance, people tend to accept hierarchies without further justification and 
authority is hardly questioned. Most Anglo and Nordic countries are characterized by small power distance scores. 
The opposite was found in Latin American countries. These nations often score high in power distance
\cite{hofstede1983national}. This pattern could explain why many tweets that question authorities (e.g., governments and companies) occur in English while they are much less present in Spanish. While this is a plausible path of reasoning, cross-cultural research has its own limitations \cite{terlutter2006globe,Tung2010}; therefore, additional work is needed to confirm or refute this explanation. 

In both the English and the Spanish dataset we found tweets from spammers and self-promoters' accounts. These accounts tend to use hashtags relevant to popular online discussions (e.g: \#Facebook, \#privacy) to promote their conversations and possibly share malicious links. These tweets account for the themes that were considered irrelevant in our analysis.
\subsection{Limitations and future work}
\label{sec:limitations}
As in any study, our research has limitations that need to be taken into consideration. We collected data through  the standard streaming Twitter API  and by using specific hashtags and keywords. Thus, we only had access to a small sample of all the tweets about the scandal. 
Nevertheless, studies have shown that this API is highly correlated with the full Twitter stream \cite{leetaru_is_nodate}. While we attempted to identify and remove bot activity from our data, and the threshold we defined lies between the recommended range [$0.43$-$0.49$] \cite{varol2017online}, our dataset contains both false positives and negatives, which could affect the results. 
Also, difficulties interpreting MEM results could occur due to the lack of contextual information such as ``information about valence and timing that is obscured in the analysis'' \cite{fitzpatrick2010beyond}. Further qualitative analysis may help with future interpretation. 

Future work includes investigating the English and Spanish corpora in order to strengthen the analysis of shared themes. This could be done, for example, by incorporating measurements of similarity between texts. Using bigrams and trigrams as inputs of the MEM analysis could help to contextualize the topic components. Also, it could be interesting to compare MEM results with other topic modeling techniques such as LDA or LDA2Vec. Furthermore, exploring the data over the time dimension should be considered because it may allow the detection of topic changes over time.

\section{Conclusions}
\label{sec:conclusions}

We proposed a methodology to conduct a cross-language study of Twitter activity about a major data privacy leakage: the Cambridge Analytica scandal. We collected tweets about it in English and Spanish during 100 days, processed them to remove tweets generated by bots, and conducted MEM to identify the key themes in each language. We conceptualized the themes and compared them. We identified five topics that were unique to only one of the languages. We also found three common topics: ``Cambridge Analytica's impact on political issues,'' ``Mark Zuckerberg's Senate hearing in the USA,'' and ``General Data Protection Regulation.'' Nevertheless, we detected dissimilarities in how these topics were formulated in each language. We proposed two plausible patterns that may explain these differences. First, reactions to a data privacy scandal reflect the users' local perspectives. Second, there is a tendency for English speakers to assign responsibilities to governments and organizations while Spanish speakers tend to attribute them to people. 

The contributions of our work and findings are twofold. First, we proposed a method to leverage social media data to investigate reactions to a data privacy scandal in two languages that cover a multi-national audience. Second, we found cross-language differences on privacy-related views that go beyond 
measuring levels of privacy concerns, as defined in widely-adopted surveys such as \cite{smith1996information}. We expect that future work can deepen the understanding of these differences through diverse methods of research. 


\section*{Acknowledgments}\label{sec:Acknowledgments}
This collaboration was possible thanks to the support of the Fulbright Program, under a 2017-18 Fulbright Fellowship award. This work was also partially funded by CONICYT Chile, under grant Conicyt-Fondecyt Iniciaci\'on 11161026. The first author acknowledges the support of the 
PIIC program 
from Universidad Técnica Federico Santa María and CONICYT-PFCHA/MagísterNacional/2019 - 22190332.

\bibliographystyle{ACM-Reference-Format}
\bibliography{referencias}

\end{document}